\documentclass{llncs}
\usepackage{epsfig}
\usepackage{amsmath}
\usepackage{amssymb}
\usepackage{algorithmic}
\usepackage{bigstrut}

\usepackage{fullpage}
\usepackage{versions}
\excludeversion{REMOVED}

\def\ColourSet{C}
\def\SolutionColourSet{C^{\star}}
\def\QueryRange{\mathcal{Q}}
\def\PointSet{\mathcal{P}}
\def\Range{R}
\def\Tree{T}
\def\NumPointsInRange{m}
\def\CandidateList{L}
\def\CandidateListUnion{L^\star}
\def\NumCandidates{k}
\def\CanonicalNodes{I}
\def\GUSFColours{U}
\def\GUSFSubsetColours{U'}
\def\WeightFunct{W}
\def\NodeDegree{f}

\def\etal{{\em et al.}}

\begin{document}
\pagestyle{plain} 

\title{Dynamic Range Majority Data
  Structures\thanks{A preliminary version of this work appeared in the
    22nd International Symposium on Algorithms and Computation (ISAAC
    2011).  This work was supported by NSERC of Canada, and NSERC
    PGS-D Scholarship, and the Canada Research Chairs Program.}}

\author{Amr Elmasry\inst{1} \and Meng He\inst{2} \and J. Ian
  Munro\inst{3} \and Patrick K. Nicholson\inst{3}}

\institute{Department of Computer Science, University of Copenhagen,
  Denmark \and Faculty of Computer Science, Dalhousie University,
  Canada \and David R. Cheriton School of Computer Science, University
  of Waterloo, Canada, \\ \email elmasry@diku.dk, mhe@cs.dal.ca,
  \{imunro, p3nichol\}@uwaterloo.ca }

\maketitle
\begin{abstract}
Given a set $\PointSet$ of $n$ coloured points on the real line, we
study the problem of answering range $\alpha$-majority (or ``heavy
hitter'') queries on $\PointSet$.  More specifically, for a query
range $\QueryRange$, we want to return each colour that is assigned to more than
an $\alpha$-fraction of the points contained in $\QueryRange$.  We
present a new data structure for answering range $\alpha$-majority
queries on a dynamic set of points, where $\alpha \in (0,1)$.  Our
data structure uses $O(n)$ space, supports queries in $O((\lg n) /
\alpha)$ time, and updates in $O((\lg n) / \alpha)$ amortized time.
If the coordinates of the points are integers, then the query time can
be improved to $O(\lg n / (\alpha \lg \lg n))$.  For constant values
of $\alpha$, this improved query time matches an existing lower bound,
for any data structure with polylogarithmic update time.  We also
generalize our data structure to handle sets of points in
$d$-dimensions, for $d \ge 2$, as well as dynamic arrays, in which
each entry is a colour.
\end{abstract}

\section{Introduction}

Many problems in computational geometry deal with point sets that have
information encoded as colours assigned to the points.  In this paper,
we design dynamic data structures for the {\em range $\alpha$-majority
  problem}, in which we want to report colours that appear
\emph{frequently} within an axis-aligned query rectangle.  This
problem is useful in database applications in which we would like to
know typical attributes of the data points in a query
range~\cite{Karpinski2008,Lai2008}.  For the one-dimensional case,
where the points represent time stamps, this problem has data mining
applications for network traffic logs, similar to those of coloured
range counting (cf.~\cite{Gagie2010}).

Formally, we are given a set, $\PointSet$, of $n$ points, where each
point $p \in \PointSet$ is assigned a colour $c$ from a set,
$\ColourSet$, of colours.  We denote the colour of $p$ as
$\text{col}(p)=c$.  We are also given a fixed parameter $\alpha \in
(0,1)$, that defines the threshold for determining whether a colour is
to be considered frequent. Our goal is to design a \emph{dynamic range
  $\alpha$-majority data structure} that can perform the following
operations:

\begin{itemize}

\item $\textsc{Query}(\QueryRange)$: We are given an axis-aligned
  hyperrectangle $\QueryRange$ as a query.  Let
  $\PointSet(\QueryRange)$ be the set $\{ p \enspace | \enspace p \in
  \PointSet \cap \QueryRange \}$, and $\PointSet(\QueryRange,c)$ be
  the set $\{ p \enspace | \enspace p \in \PointSet(\QueryRange),
  \text{col}(p) = c \}$.  The answer to the query $\QueryRange$ is the
  set of colours $\SolutionColourSet$ such that for each colour $c \in
  \SolutionColourSet$, $|\PointSet(\QueryRange,c)| > \alpha
  |\PointSet(\QueryRange)|$, and for all $c \not \in
  \SolutionColourSet$, $|\PointSet(\QueryRange,c)| \le \alpha
  |\PointSet(\QueryRange)|$.  We refer to a colour $c \in
  \SolutionColourSet$ as an \emph{$\alpha$-majority} for
  $\QueryRange$, and this type of query as an \emph{$\alpha$-majority
    query}.  When $\alpha = 1/2$, the problem is to identify
  the majority colour in $\QueryRange$, if such a colour exists.

\item $\textsc{Insert}(p,c)$: Insert a point $p$ with colour $c$ into
  $\PointSet$.

\item $\textsc{Delete}(p)$: Remove the point $p$ from $\PointSet$.

\end{itemize}

\subsection{Previous Work\label{sec:prevresults}}

\paragraph{Static and Dynamic Range $\alpha$-Majority:}  

In all of the following results, unless mentioned otherwise, the
threshold $\alpha \in (0,1)$ is fixed at construction time, rather
than specified for each query individually.

Karpinski and Nekrich~\cite{Karpinski2008} studied the problem of
answering range $\alpha$-majority queries, which they call
\emph{coloured $\alpha$-domination} queries. In the static case, they
gave an $O(n/\alpha)$ space data structure that supports
one-dimensional queries in $O((\lg n \lg \lg n) / \alpha)$
time\footnote{We use $\lg n$ to denote $\lceil \log_2 n \rceil$. },
and an $O((n \lg \lg n) / \alpha)$ space data structure that supports
queries in $O((\lg n) / \alpha)$ time.  In the dynamic case, they gave
an $O(n/\alpha)$ space data structure for one-dimensional queries that
supports queries and insertions in $O((\lg^2 n) / \alpha)$ time, and
deletions in $O((\lg^2 n) / \alpha)$ amortized time.  They also gave
an alternative $O( (n \lg n) / \alpha)$ space data structure that
supports queries and insertions in $O((\lg n) / \alpha)$ time, and
deletions in $O((\lg n) / \alpha)$ amortized time. For points in
$d$-dimensions, for constant $d \ge 2$, they gave a static $O((n
\lg^{d-1} n) / \alpha)$ space data structure that supports queries in
$O((\lg^{d} n) / \alpha)$ time, as well as a dynamic $O((n \lg^{d-1}
n) / \alpha)$ space data structure that supports queries and
insertions in $O((\lg^{d+1} n) / \alpha)$ time, and deletions in
$O((\lg^{d+1} n) / \alpha)$ amortized time.

Durocher~\etal~\cite{DHMNS2011} described a static
$O(n(\lg(1/\alpha)+1))$ space data structure that answers range
$\alpha$-majority queries in an array in $O(1/\alpha)$ time.  This
data structure is based on the idea that it is possible to produce a
short list of candidate $\alpha$-majorities for any query, and then
efficiently verify the frequencies of these candidates using succinct
data structures.  In a later version of the same
paper~\cite{DHMNS2012}, they described how to extend their technique
to $d$-dimensions for constant $d \ge 2$, resulting in an $O(n
\lg^{d-1} n)$ space data structure that supports range
$\alpha$-majority queries in $O(\lg^{d} n/\alpha)$
time. Gagie~\etal~\cite{GHMN2011} improved the static one-dimensional
result to $O(n(\min(\lg(1/\alpha),H)+1))$ space, where $H \le \lg n$
is the 0th-order empirical entropy of the sequence stored in the
array.  The same authors also described how to improve the query time
to $O(1/\beta)$, when asked for the $\beta$-majorities in a query
range, for any $\beta \ge \alpha$ specified at query time.  Recently,
for the two-dimensional static case, Wilkinson~\cite{W12} presented an
improved data structure that occupies $O(n\lg^{\varepsilon}n
\lg(1/\alpha))$ space, for any constant $\varepsilon > 0$, and can
answer queries in $O(\lg n / \alpha)$ time.

\paragraph{Approximate Versions of the Problem:} 
Researchers have also examined an approximate version of the range
$\alpha$-majority problem, in which the solution must contain all the
$\alpha$-majorities in a query range, but can also contain some false
positives.  Lai~\etal~\cite{Lai2008} studied the dynamic problem,
using the term \emph{heavy-colours} instead of
$\alpha$-majorities. They presented a dynamic data structure based on
sketching, which provides an approximate solution with probabilistic
guarantees for constant values of $\alpha$.  For one dimension their
data structure uses $O(hn)$ space and supports queries and updates in
$O(h\lg n)$ time, where the parameter $h = O(\frac{\lg
  |\ColourSet|}{\varepsilon}\lg (\frac{\lg |\ColourSet|}{\alpha
  \delta}))$ depends on the threshold $\alpha$, the approximation
factor $\varepsilon$, the total number of colours $|\ColourSet|$, and
the probability of failure $\delta$.  They also noted that the space
can be reduced to $O(n)$, at the cost of increasing the query time to
$O(h\lg n + h^2)$.  Thus, for constant values of $\varepsilon$,
$\delta$, and $\alpha$, their data structure uses $O(n)$ space and has
$O((\lg n\lg \lg n)^2)$ query and update time in the worst case when
$\lg m = \Omega(\lg n)$.

Another approximate data structure based on sketching was proposed by
Wei and Yi~\cite{Wei11}.  Their data structure uses linear space,
answers queries in $O(\lg n + 1/\varepsilon)$ time, and may return
false positives with relative frequency between $\alpha - \varepsilon$
and $\alpha$.  The cost of updates is $O(\mu\lg n \lg
(1/\varepsilon))$ amortized time, where $\mu$ is the cost of updating
the sketches.  We note that this result was obtained independently of
ours, and that both our techniques and the main technique they
develop, called \emph{exponential decomposability}, are similar.  By
combining Theorem 4 of their paper with standard range counting data
structures, it is not difficult to get a data structure that occupies
linear space, answers queries in $O(\lg n/\alpha)$ time, and supports
updates in $O((\lg n \lg (1/\alpha))/\alpha)$ amortized time for the
non-approximate version of the problem that we study.  However, we
slightly improve this update time, and also generalize our data
structures to higher dimensions, whereas their structure is part of a
more general framework that supports other kinds of aggregate queries.

\paragraph{Lower Bounds:} 
The partial sum problem for threshold functions~\cite{Husfeldt2003}
captures the essence of the dynamic range $\alpha$-majority problem:
maintain $n$ bits $x_1,...,x_n$ subject to updates and \emph{threshold
  queries}. An update consists of flipping the bit at a specified
index. The answer to query $\texttt{threshold}(i)$ is ``yes'' if and
only if $\sum_{j=1}^{i} x_j \ge f(i)$, where $f(i)$ is an integer
function such that $f(i) \in \{0,...,\lceil i/2 \rceil \}$.  Husfeldt
and Rauhe proved a lower bound~\cite{Husfeldt2003} on the query time
$t_q$ for a data structure that can answer threshold queries with
update time $t_u$.

Any data structure for dynamic $\alpha$-majority can be used to solve
the partial sum problem for threshold functions.  In particular, we
can treat the problem as involving $n$ points with integer coordinates
$1, ..., n$, with each point having one of two colours.  A flip
operation can be implemented as a deletion followed by an insertion.
Thus, we can state their lower bound in terms of our problem, denoting
the cell size of our machine as $w$:

\begin{lemma}[Follows from~\cite{Husfeldt2003}, Prop. 4] \label{lem:lb}
Let $t_u$ and $t_q$ denote the update and query times, respectively,
for any dynamic $\alpha$-majority data structure.  Then,
\begin{equation}
t_q = \Omega \left( \frac{\lg (\min \{ \alpha n, (1-\alpha)n \} )}{\lg(t_u w \lg (\min \{ \alpha n, (1-\alpha)n \}))}\right) \enspace . \notag
\end{equation}
\end{lemma}

This bound suggests that, for constant values of $\alpha$ and word
size $\Theta(\lg n)$ bits, $O(\lg n/\lg \lg n)$ query time for integer
point sets is optimal for any data structure with polylogarithmic
update time.

\paragraph{Other Related Work:} 

Finally, several other results exist for finding $\alpha$-majorities
in the streaming model, typically referred to as \emph{heavy
  hitters}~\cite{boyer1991,Demaine2002,Karp2003,Misra1982}.  De~Berg
and Haverkort~\cite{DeBerg2003} studied a similar problem of reporting
$\tau$-\emph{significant} colours.  For this problem, the goal is to
output all colours $c$ such that at least a $\tau$-fraction of
\emph{all} of the points with colour $c$ lie in the axis-aligned query
rectangle.  More broadly, there are other data structure problems that
deal with coloured points.  In {\em coloured range reporting
  problems}, we are interested in reporting the set of distinct
colours assigned to the points contained in an axis-aligned rectangle.
Similarly, in {\em the coloured range counting problem} we are
interested in returning the number of such distinct colours.
Gupta~\etal~\cite{Gupta1995}, Bozanis~\etal~\cite{Bozanis1995}, and,
more recently, Gagie et al.~\cite{GNP2010} and Gagie and
K{\"a}rkk{\"a}inen~\cite{Gagie2010} studied these problems and
presented several interesting results.

\subsection{Our Results\label{sec:result}}

In this paper we present new data structures for the dynamic range
$\alpha$-majority problem in the word-RAM model with word size $\Omega
( \lg n)$, where $n$ is the number of points in the set $\PointSet$,
and $\alpha \in (0,1)$. Our results are summarized and compared to the
previous best results in Table~\ref{tab:results}.  The \emph{input}
column indicates the type of data we are considering.  We use
\emph{points} to denote a set of points on a line with real-valued
coordinates that we can compare in constant time, \emph{integers} to
denote a set of points on a line with word sized integer coordinates,
and \emph{array} to denote that the input is to be considered a
dynamic array, where the positions of the points are in rank space.

\begin{table}[width=\textwidth]
\centering
\begin{tabular}{|l|l|l|l|l|l|l|}
\hline 
Source & Input &  Space & Query & Insert & Delete  \bigstrut \\ 
\hline
\cite[Thm. 3]{Karpinski2008} & points & $O(\frac{n}{\alpha})$ & $O(\frac{\lg^2 n}{\alpha})$ & $O(\frac{\lg^2 n}{\alpha})$ & $O(\frac{\lg^2 n}{\alpha})$* \bigstrut \\ 
\cite[Thm. 3]{Karpinski2008} & points & $O(\frac{n\lg n}{\alpha})$ & $O(\frac{\lg n}{\alpha})$ & $O(\frac{\lg n}{\alpha})$ & $O(\frac{\lg n}{\alpha})$* \bigstrut \\ 
New & points & $O(n)$ & $O(\frac{\lg n}{\alpha})$ & $O(\frac{\lg n}{\alpha})$* & $O(\frac{\lg n}{\alpha})$*  \bigstrut \\ 
New & integers & $O(n)$ & $O(\frac{\lg n}{\alpha \lg \lg n})$& $O(\frac{\lg n}{\alpha})$* &  $O(\frac{\lg n}{\alpha})$*  \bigstrut \\ 
New & array & $O(n)$ & $O(\frac{\lg n}{\alpha \lg \lg n})$ & $O(\frac{\lg^3 n}{\alpha \lg \lg n})$* & $O(\frac{\lg n}{\alpha})$* \bigstrut \\ 
\hline
\end{tabular}
\caption{\label{tab:results}Comparison of the results in this paper to
  the previous best results.  For the entries marked with ``*'' the
  running times are amortized.}
\end{table}

Our results improve upon previous results in several ways.  Most
noticeably, all our data structures require linear space.  In order to
provide fast query and update times for our linear space structures,
we prove several interesting properties of $\alpha$-majority colours.
We note that the lower bound from Lemma~\ref{lem:lb} implies that, for
constant values of $\alpha$, an $O(\lg n/\lg \lg n)$ query time for
integer point sets is optimal for any data structure with
polylogarithmic update time, when the word size $w = \Theta(\lg n)$.
Our data structure for points on a line with integer coordinates
achieves this optimal query time.

Our data structures can also be generalized to handle $d$-dimensional
points, improving upon previous results in the dynamic
case~\cite{Karpinski2008}. For $d \ge 2$, our data structure occupies
$O(n \lg^{d-1} n)$ space, answers range $\alpha$-majority queries in
$O((\lg^{d}n)/\alpha)$ time, and supports updates in
$O((\lg^{d}n)/\alpha)$ amortized time.

\paragraph{Road Map:} 
In Section~\ref{sec:1dds} we present a dynamic range $\alpha$-majority
data structure for points in one dimension.  In Section
\ref{sec:intkeys} we show how to speed up the query time of our data
structure in the case where the points have integer coordinates.  In
Section~\ref{sec:higher-d} we generalize our one dimensional data
structures to higher dimensions.  Finally, in
Section~\ref{sec:array} we present our data structure for dynamic
arrays.

\paragraph{Assumptions About Colours:}
In the following sections, we assume that we can compare colours in
constant time.  In order to support a dynamic set of colours, we
employ the techniques described by Gupta~\etal~\cite{Gupta1995}.
These techniques allow us to maintain a mapping from the set of
colours to integers in the range $[1,2n]$, where $n$ is the number of
points \emph{currently} in our data structure.  This allows us to
index into an array using a colour in constant time.

For the dynamic problems discussed, this mapping is maintained using a
method similar to global rebuilding to ensure that the integer
identifiers of the colours do not grow too large~\cite[Section
  2.3]{Gupta1995}.  When a coloured point is inserted, we must first
determine whether we have already assigned an integer to that colour.
By storing the set of known colours in a balanced binary search tree,
this can be checked in $O(\lg |\ColourSet|)$ time; recall that
$|\ColourSet|$ is the number of distinct colours currently assigned to
points in our data structure.  Since $|\ColourSet| \le n$, this cost
is absorbed by update time of our data structure; see
Table~\ref{tab:results}.  Therefore, from this point on, we assume
that we are dealing with integers in the range $[1,2n]$ when we
discuss colours.

\section{\label{sec:1dds}Dynamic Data Structures in One Dimension}

In one-dimension we can interchange the notion of points and
$x$-coordinates in $\PointSet$, since they are equivalent.  Depending
on the context we may use either term.  Our basic data structure, like
that of Karpinski and Nekrich~\cite{Karpinski2008}, is a modified
weight balanced B-tree~\cite{av2003}.  However, we prove several
interesting combinatorial properties of $\alpha$-majorities in order
to provide more efficient support for queries and updates.

Let $\Tree$ be a weight-balanced B-tree with branching parameter $8$
and leaf parameter $1$ such that each leaf represents an
$x$-coordinate in $\PointSet$.  From left to right the leaves are
sorted in ascending order of the $x$-coordinate that they
represent. Let $\Tree(u)$ be the subtree rooted at node $u$. Each
internal node $u$ in the tree represents a range $\Range(u) =
[x_{\min},x_{\max}]$, where $x_{\min}$ is the $x$-coordinate represented by
the leftmost leaf in $\Tree(u)$, and $x_{\max}$ is the $x$-coordinate
represented by the rightmost leaf in $\Tree(u)$. We number the levels of
the tree $0, 1, ..., \Theta(\lg n)$ from top to bottom.  If a node is
$h$ levels above the leaf level, we say that this node is of {\em
  height} $h$.  By the properties of weight-balanced B-trees, the
range represented by an internal node of height $h$ contains at least
$8^h / 2$ (except the root) and at most $2 \times 8^h$ points, and the
degree of each internal node is at least $2$ and at most $32$.
 
\subsection{Supporting Queries\label{sec:supquery}}
Given a query $\QueryRange' = [x_a',x_b']$, we perform a top-down
traversal on $\Tree$ to map $\QueryRange'$ to the range $\QueryRange =
[x_a,x_b]$, where $x_a$ and $x_b$ are the points in $\PointSet$ with
$x$-coordinates that are the successor and the predecessor of $x_a'$
and $x_b'$, respectively.  We call the query range $\QueryRange$ {\em
  general} if $\QueryRange$ is not represented by a single node of
$\Tree$.  We first define the notion of {\em representing} a general
query range by a set of nodes:

\begin{definition}
Given a general query range $\QueryRange = [x_a, x_b]$, $\QueryRange$
induces a set, $\CanonicalNodes$, of nodes in the tree $\Tree$,
satisfying the following two conditions.  
\begin{enumerate}
\item The range represented by the parent of each node in
  $\CanonicalNodes$ is not entirely contained in $\QueryRange$.  
\item For all $p \in \PointSet \cap \QueryRange$, there exists some
  node $u \in \CanonicalNodes$ with $p \in \Range(u)$.  
\end{enumerate}
We say that $\CanonicalNodes$ is the set of nodes in the tree $\Tree$
\emph{representing} $\QueryRange$.
\end{definition}

For each node $u \in \Tree$, we keep a list, $\CandidateList(u)$, of
$\NumCandidates$ \emph{candidate} colours, i.e., the $\NumCandidates$
most frequent colours in the range $\Range(u)$ represented by $u$,
breaking ties arbitrarily.  Later, we will fix a value for
$\NumCandidates$.  Let $\CandidateListUnion = \cup_{u \in
  \CanonicalNodes} \CandidateList(u)$, i.e., the union of all the
candidate lists among the nodes representing the query range
$\QueryRange$.  For each colour $c \in \ColourSet$, we keep a separate
range counting data structure, $F_c$, containing all points $p \in
\PointSet$ with colour $c$, and also a range counting data structure,
$F$, containing all of the points in $\PointSet$.  Let
$\NumPointsInRange$ be the total number of points in the range $[x_a,
  x_b]$, which can be determined by querying $F$. For each $c \in
\CandidateListUnion$, we query $F_c$ with the range $[x_a,x_b]$
letting $\texttt{occ}$ be the result. If $\texttt{occ} > \alpha
\NumPointsInRange$, then we report that $c \in \SolutionColourSet$.

It is clear that $\CanonicalNodes$ contains at most $\Theta(\lg n)$
nodes.  Furthermore, if a colour $c$ is an $\alpha$-majority for
$\QueryRange$, then it must be an $\alpha$-majority for at least one
of the ranges in $\CanonicalNodes$ \cite[Observation
  1]{Karpinski2008}. If we set $\NumCandidates = \lceil 1/\alpha
\rceil$ and store $\lceil 1/\alpha \rceil$ colours in each internal
node as candidate colours, then, by the procedure just described, we
will perform a range counting query on $\Theta((\lg n) / \alpha)$
colours.  If we use balanced search trees for our range counting data
structures, then this takes $\Theta((\lg ^2 n) / \alpha)$ time
overall.  However, in the sequel we show how to improve this query
time by exploiting the fact that the nodes in $\CanonicalNodes$ that
are closer to the root of $\Tree$ contain more points in the ranges
that they represent.

We shall prove useful properties of a general query range
$\QueryRange$ and the set, $\CanonicalNodes$, of nodes representing it
in Lemmas~\ref{lem:height},~\ref{lem:notcandidate},~\ref{lem:top},
and~\ref{lem:parameters}. In these lemmas, $\NumPointsInRange$ denotes
the number of points in $\QueryRange$, and $i_1, i_2, ...$ denote the
distinct values of the heights of the nodes in $\CanonicalNodes$,
where $i_1 > i_2 > ... \ge 0$.  We first give an upper bound on the
number of points contained in the ranges represented by the nodes of
$\CanonicalNodes$ of a given height:

\begin{lemma}~\label{lem:height}
The total number of points in the ranges represented by all the nodes
in $\CanonicalNodes$ of height $i_j$ is less than $\NumPointsInRange
\times \min(1,31 \times 8^{1-j})$.
\end{lemma}

\begin{proof}
Since $\QueryRange$ is general and contains at least one node of
height $i_1$, $\NumPointsInRange$ is greater than the minimum number
of points that can be contained in a node of height $i_1$, which is
$8^{i_1} / 2$.  The nodes of $\CanonicalNodes$ whose height is $i_j$,
$j \neq 1$, are siblings and must have at least one sibling that is
not in $\CanonicalNodes$.  The number of points contained in the
interval represented by this sibling is at least $8^{i_j}/2$.
Therefore, the number, $\NumPointsInRange_j$, of points in the ranges
represented by the nodes of $\CanonicalNodes$ at level $i_j$ is at
most $2 \times 8^{i_j+1} - 8^{i_j}/2 = (31/2) \times 8^{i_j} $.  Thus,
$\NumPointsInRange_j / \NumPointsInRange < 31 \times 8^{i_j-i_1} < 31
\times 8^{1-j}$. \qed
\end{proof}

We next use the above lemma to bound the number of points whose colours
are not among the candidate colours stored in the corresponding nodes
in $\CanonicalNodes$.

\begin{lemma}\label{lem:notcandidate}
Suppose we are given a node $v \in \CanonicalNodes$ of height $i_j$
and a colour $c$.  Let $n_v^{(c)}$ denote the number of points with
colour $c$ in $\Range(v)$, the range covered by $v$, if $c$ is not
among the first $\NumCandidates_j = \lceil \NumCandidates/2^{j-1}
\rceil$ most frequent candidate colours in the candidacy list of $v$,
and $n_v^{(c)} = 0$ otherwise. Then $\sum_{v \in \CanonicalNodes}
n_v^{(c)} < \frac{5.59 \NumPointsInRange}{\NumCandidates+1}$.
\end{lemma}

\begin{proof}
If $c$ is not among the first $\NumCandidates_j$ candidate colours
stored in $v$, then the number of points with colour $c$ in
$\Range(v)$ is at most $1/(\NumCandidates_j+1)$ times the number of
points in $\Range(v)$. Thus,

\begin{eqnarray*}
\sum_{v \in \CanonicalNodes} n_v^{(c)} & < & \sum_{j=1}^2 \frac{\NumPointsInRange}{\NumCandidates_j +1} +  \sum_{j \geq 3} \frac{\left(31 \times 8^{1-j}\right)\NumPointsInRange }{\NumCandidates_j +1}   \\
& < & \frac{\NumPointsInRange}{\NumCandidates +1} \left(1 + 2 + 31 \left(\frac{2^2}{8^2} + \frac{2^3}{8^3}+ \dots\right)\right)\\
& < & \frac{5.59\NumPointsInRange}{\NumCandidates+1} 
\end{eqnarray*}
\qed
\end{proof}

We next consider the nodes in $\CanonicalNodes$ that are closer to the
leaf levels.  Let $\CanonicalNodes_t$ denote the nodes in
$\CanonicalNodes$ that are at one of the top $t =
\lceil\frac{\lg(\frac{1}{\alpha})}{3} + 2.05\rceil$--- not necessarily
consecutive--- levels of the nodes in $\CanonicalNodes$.  We prove the
following property:

\begin{lemma}\label{lem:top}
The number of points contained in the ranges represented by the nodes
in $\CanonicalNodes \setminus \CanonicalNodes_t$ is less than $\alpha
\NumPointsInRange /2$.
\end{lemma}

\begin{proof}
By Lemma~\ref{lem:height}, the number of points contained in the
ranges represented by the nodes in $\CanonicalNodes \setminus
\CanonicalNodes_t$ is less than:
\begin{eqnarray}
31\NumPointsInRange \sum_{j \ge t+1}8^{1-j} &<& 31 \NumPointsInRange \left(\frac{1}{8^{t}}+\frac{1}{8^{t+1}} + ...\right) \nonumber\\
&<& 31\NumPointsInRange \left(\frac{8}{7} \times \frac{1}{8^{t}}\right) \nonumber
\end{eqnarray}

Since $t \ge \frac{\lg(\frac{1}{\alpha})}{3} + 2.05$, the above value
is less than $\alpha \NumPointsInRange /2$.  \qed
\end{proof}

With the above lemmas, we can choose an appropriate value for
$\NumCandidates$ to guarantee the following property that is critical
to achieve improved query time:

\begin{lemma}~\label{lem:parameters}
When $\NumCandidates = \lceil \frac{11.18}{\alpha} \rceil - 1$, any
$\alpha$-majority colour, $c$, of the query range $\QueryRange$ is
among the union of the first $\lceil \NumCandidates/2^{j-1} \rceil$
candidates stored in each node of height $i_j$ representing a range in
$\CanonicalNodes_t$.
\end{lemma}

\begin{proof}
The total number of points with colour $c$ in the ranges represented
by the nodes in $\CanonicalNodes \setminus \CanonicalNodes_t$ is less
than $\alpha \NumPointsInRange /2$ by Lemma~\ref{lem:top}.  By
Lemma~\ref{lem:notcandidate} and our choice for the value of
$\NumCandidates$, less than $\alpha \NumPointsInRange /2$ points in the ranges
represented by the nodes in $\CanonicalNodes_t$ for which $c$ is not a
candidate can have colour $c$.  The lemma thus follows.  \qed
\end{proof}

For each node $v \in \Tree$, we keep a semi-ordered list of the
$\NumCandidates$ candidate colours in the range $\Range(v)$
represented by $v$. The order on the colours for any candidacy list is
maintained such that the most frequent $\lceil
\NumCandidates/2^{j-1}\rceil$ colours come first, for all
$j=2,3,\dots$, arbitrarily ordered within their positions.  Note that
such a semi-ordering can be obtained in $O(\NumCandidates)$ time by
repeated median queries.  That is, by using a linear time median
finding algorithm~\cite{BFPRT1973}, we can partition the list so that
the first half of the list contains the $\NumCandidates/2$ most
frequent colours, and then recurse on the first half of the list until
the list has $1$ element.  In total, this takes $O(\NumCandidates +
\NumCandidates/2 + ... \NumCandidates/4) = O(\NumCandidates)$ time.

By setting $\NumCandidates = \lceil 11.18 / \alpha \rceil - 1$,
Lemma~\ref{lem:parameters} implies that the colours that we have
checked are the only possible $\alpha$-majority colours for the query.
Furthermore, Lemma~\ref{lem:top} implies that we need only check the
nodes on the top $O(\lg(1/\alpha))$ levels in $\CanonicalNodes$.  Let
$\CanonicalNodes_t$ denote the set of nodes in these levels. We
present the following lemma:

\begin{lemma}
\label{lem:query}
The data structures described in this section occupy $O(n)$ words, and
can be used to answer a range $\alpha$-majority query in $O((\lg n)
/\alpha)$ time with the help of an additional array of size $2n$.
\end{lemma}

\begin{proof}
To support $\alpha$-majority queries, we only consider the nontrivial
case in which the query range $\QueryRange$ is general.  By
Lemma~\ref{lem:parameters}, the $\alpha$-majorities can be found by
examining the first $\lceil \NumCandidates/2^{j-1} \rceil$ candidate
colours stored in each node representing a range in
$\CanonicalNodes_t$.  Thus, there are at most $O(\lceil
\frac{1}{\alpha} \rceil + \lceil \frac{1}{2\alpha} \rceil + \lceil
\frac{1}{4\alpha}\rceil + ... + \lceil \frac{1}{2^{t-1} \alpha}\rceil)
= O(\frac{1}{\alpha})$ relevant colours to check.  Let
$\CandidateList_t$ denote the set of these colours.  For each $c \in
\CandidateList_t$ we query our range counting data structures $F_c$
and $F$ in $\Theta(\lg n)$ time to determine whether $c$ is an
$\alpha$-majority.  Thus, the overall query time is $O((\lg
n)/\alpha)$.

There are $\Theta(n)$ nodes in the weight-balanced B-tree. Therefore,
one would expect the space to be $\Theta(n/\alpha)$ words, since each
node stores $\Theta(1/\alpha)$ colours.  We use a pruning technique on
the lower levels of the tree in order to reduce the space to $O(n)$
words overall.  If a node $u$ covers less than $1/\alpha$ points, then
we need not store $\CandidateList(u)$, since every colour in
$\Tree(u)$ is an $\alpha$-majority for $\Range(u)$.  Instead, during a
query, we can traverse the leaves of $\Tree(u)$ in order to determine
the unique colours.  To make this efficient, we require an array $D$
of size $2n$ integers to count the frequencies of the colours in
$\Range(u)$.  As mentioned in Section~\ref{sec:result}, we can map a
colour into an index of the array $D$, which allows us to increment a
frequency counter in $O(1)$ time.  Thus, we can extract the unique
colours in $\Range(u)$ in $O(|\Tree(u)|) = O(\frac{1}{\alpha})$ time.
The number of tree nodes whose subtrees have at least $1/\alpha$
leaves is $O(n \alpha)$.  Thus, we store $O(\NumCandidates) =
O(1/\alpha)$ words in $O(n \alpha)$ nodes, and the total space used by
our B-tree $\Tree$ is $O(n)$ words.  The only other data structures we
make use of are the array $D$ and the range counting data structures
$F$ and $F_c$ for each $c\in \ColourSet$, and together these occupy
$O(n)$ words.  \qed
\end{proof}

\subsection{Supporting Updates}

We next establish how much time is required to maintain the list
$\CandidateList(v)$ in node $v$ under insertions and deletions.  We
begin by observing that it is not possible to lazily maintain the list
of the top $\NumCandidates=\lceil \frac{11.18}{\alpha} \rceil -1$ most
frequent colours in each range: many of these colours could have low
frequencies, and the list $\CandidateList(v)$ would have to be rebuilt
after very few insertions or deletions.  To circumvent this problem,
we relax our requirements on what is stored in $\CandidateList(v)$,
only guaranteeing that \emph{all} of the $\beta$-majorities of the
range $\Range(v)$ must be present in $\CandidateList(v)$, where $\beta
= \lceil\frac{11.18}{\alpha}\rceil^{-1}$.  With this alteration, we
can still make use of the lemmas from the previous section, since they
depend only on the fact that there are no colours $c \not \in
\CandidateList(v)$ with frequency greater than $\beta |\Tree(v)|$.
The issue now is how to maintain the $\beta$-majorities of $\Range(v)$
during insertions and deletions of colours.

Karpinski and Nekrich noted that if we store the
$(\beta/2)$-majorities for each node $v$ in $\Tree$, then it is only
after $|\Tree(v)|\beta/2$ deletions that we must rebuild
$\CandidateList(v)$~\cite{Karpinski2008}.  For the case of insertions
and deletions, their data structure performs a range counting query at
each node $v$ along the path from the root of $\Tree$ to the leaf
representing the inserted or deleted colour $c$.  This counting query
is used to determine if the colour $c$ should be added to, or removed
from, the list $\CandidateList(v)$.

In contrast, our strategy is to be lazy during insertions and
deletions, waiting as long as possible before recomputing
$\CandidateList(v)$, and to avoid performing range counting queries
for each node in the update path. We provide a tighter analysis (to
constant factors) of how many insertions and deletions can occur
before the list $\CandidateList(v)$ is to be rebuilt.  One caveat is
that the results in this section only apply when $\beta \in
(0,\frac{1}{2}]$.  However, since $\alpha <1$, our choice of $\beta$
  satisfies this condition.

We use $\mathbb{Z}^{*}$ to denote $\mathbb{Z^+} \cup \{ 0 \}$.  The
following lemma is used to show a lower bound on the number of update
operations (insertions and deletions) that can occur before a list
needs to be recomputed:

\begin{lemma}
\label{lem:gammafunc}
Let $\Gamma (\ell, j, \beta) = \min_{n_i \in \mathbb{Z}^*, n_d \in
  \mathbb{Z}^*} \left\{ n_i + n_d \enspace \Big| \enspace \frac{j +
  n_i}{\ell + n_i - n_d} > \beta \right\}$ where $\ell \in
\mathbb{Z}^+$, $j \in \mathbb{Z}^*$, $j < \ell$, and $\beta \in
(0,\frac{1}{2}]$.  If $\ell \ge 2j + 1$, then $\Gamma (\ell, j, \beta)
  \ge \frac{\beta \ell - j}{1 - \beta}$.
\end{lemma}

\begin{proof}
Observe that $\frac{j+1}{\ell+1} \ge \frac{j}{\ell - 1}$ if and only
if $\ell \ge 2j + 1$.  This implies that increasing $n_i$ rather than
$n_d$ by the same amount increases the value of the ratio $\frac{j +
  n_i}{\ell + n_i - n_d}$ by a greater amount when $\ell + n_i - n_d
\ge 2(j + n_i)+1$. Thus, we have
\begin{eqnarray}
\Gamma(\ell, j, \beta) = \min_{n_i \in Z^*} \left\{n_i \enspace \big|\enspace \frac{m+n_i}{\ell+n_i} > \beta\right\} \nonumber \enspace,
\end{eqnarray}
\noindent
if $\ell +i \ge 2(j+i) + 1$ for $1 \le i < \Gamma(\ell,j,\beta)$.
Also observe that $\frac{j + n_i}{\ell + n_i} > \beta$ implies $n_i >
\frac{\beta \ell - j }{1-\beta}$. All that remains is to show that for
$\beta \in (0,\frac{1}{2}]$ the constraint $\ell +i \ge 2(j+i) + 1$ is
  satisfied for $1 \le i < \Gamma(\ell,j,\beta)$.  To show this, we
  observe that if $i < \ell - 2j$, then $\ell + i \ge 2(j+i)+1$.
  Thus, the constraint is satisfied if $\Gamma(\ell,j,\beta) \le \ell
  - 2j$.  Since $\frac{\beta \ell - j }{1-\beta} \le \ell - 2j$ for
  all $\beta \in (0, \frac{1}{2}]$, we get the desired bound. \qed
\end{proof}

We can think of the variables $n_i$ and $n_d$ as the numbers of
insertions and deletions into our data structure.  Thus,
$\Gamma(\ell,j,\beta)$ represents the number of updates that can occur
in a range containing $\ell$ points before a colour $c$ with $j$
occurrences can possibly become a $\beta$-majority.  We next prove the
following lemma:

\begin{lemma}
\label{lem:rebuildlists}
Suppose the list $\CandidateList(v)$ for node $v$ contains the $\lceil
\frac{1-\beta + \sqrt{1-\beta}}{\beta} \rceil$ most frequent colours
in the range $\Range(v)$, breaking ties arbitrarily.  For $\beta \in
(0,\frac{1}{2}]$, this value is upper bounded by $\lceil
  \frac{2}{\beta} \rceil$. Let $\ell$ be the number of points
  contained in $\Range(v)$.  Only after $\lceil
  \frac{\beta\ell}{\sqrt{1-\beta}(1+\sqrt{1-\beta})} \rceil \ge \lceil
  \frac{\beta\ell}{2} \rceil$ insertions or deletions into $\Tree(v)$
  can a colour $c \not \in \CandidateList(v)$ possibly become a
  $\beta$-majority for the range spanned by node $v$.
\end{lemma}

\begin{proof}
Since we keep in $\CandidateList(v)$ the $\NumCandidates$ most
frequently appearing colours in the range $\Range(v)$, any colour not
in $\CandidateList(v)$ can appear at most
$\frac{\ell}{\NumCandidates+1}$ times.  We apply Lemma
\ref{lem:gammafunc}, noting that it exactly describes the number of
insertions or deletions required to cause a colour with frequency $m$
to become a $\beta$-majority in a range containing $\ell$ points.
Thus, we get that $\Gamma(\ell, \frac{\ell}{\NumCandidates+1}, \beta)
\ge \frac{\beta \ell - \ell / (\NumCandidates+1)}{1 - \beta}$.  We
want to maximize the ratio $\Gamma(\ell,
\frac{\ell}{\NumCandidates+1}, \beta)/\NumCandidates$, which gives us
the maximum number of updates before rebuilding $\CandidateList(v)$
per colour stored in $v$.  If $h(\NumCandidates) = \frac{\beta \ell
  -\ell/(\NumCandidates+1)}{(1-\beta)\NumCandidates}$, then the
derivative $h'(\NumCandidates) = \frac{(-2\NumCandidates+\beta
  \NumCandidates^2+2\beta
  \NumCandidates+\beta-1)\ell}{(\NumCandidates+1)^2(-1+\beta)\NumCandidates^2}$,
which has zeros at $\NumCandidates = \{
\frac{1-\beta+\sqrt{1-\beta}}{\beta},\frac{1-\beta-\sqrt{1-\beta}}{\beta}
\}$.  The relevant zero, which maximizes $h(\NumCandidates)$ is
$\NumCandidates = \frac{1-\beta+\sqrt{1-\beta}}{\beta}$.  Substituting
this in as $\NumCandidates$ into $\frac{\beta \ell -
  \ell/(\NumCandidates+1)}{1-\beta}$, we get that
$\frac{\beta\ell}{\sqrt{1-\beta}(1+\sqrt{1-\beta})}$ updates are
required before a colour $c \not \in \CandidateList(v)$ can become a
$\beta$-majority for the range spanned by node $v$. \qed
\end{proof}

By Lemma~\ref{lem:rebuildlists}, our lazy updating scheme only
requires each list $\CandidateList(v)$ to have size $\lceil
\frac{1-\beta + \sqrt{1-\beta}}{\beta} \rceil = O(1/\alpha)$.  This
leads to the following theorem:

\begin{theorem}\label{thm:dyn-1d-alpha-maj}
Given a set $\PointSet$ of $n$ points in one dimension and a fixed
$\alpha \in (0,1)$, there is an $O(n)$ space data structure that
supports range $\alpha$-majority queries on $\PointSet$ in $O((\lg n)
/\alpha)$ time, and insertions and deletions in $O((\lg n) / \alpha)$
amortized time.
\end{theorem}

\begin{proof}
Query time follows from Lemma~\ref{lem:query}. In order to get the
desired space, we combine Lemmas~\ref{lem:query} and
\ref{lem:rebuildlists}, implying that each list $\CandidateList(v)$
contains $O(1/\alpha)$ colours.  This allows us to use the same
pruning technique described in Lemma~\ref{lem:query} in order to
reduce the space to $O(n)$.

When an update occurs, we follow the path from the root of $\Tree$ to
the updated node $u$.  Suppose, without loss of generality, the update
is an insertion of a point of colour $c$.  For each vertex $v$ on the
path, if $v$ contains a list $\CandidateList(v)$, we check whether $c$
is in $\CandidateList(v)$.  If it is, then we increment the count of
colour $c$. This takes $O(1/\alpha)$ time.  We also increment the
counter for $v$ that keeps track of the number of updates into
$\Tree(v)$ that have occurred since $\CandidateList(v)$ was rebuilt.
Thus, modifying the lists and counters along the path requires $O((\lg
n)/\alpha)$ time in the worst case.

We next look at the costs of maintaining the lists
$\CandidateList(v)$.  The list $\CandidateList(v)$ can be rebuilt in
$O(|\Tree(v)|)$ time, using the array $D$.  Note that $D$ can be
maintained under updates using the same scheme described in
Section~\ref{sec:result}.  First, we use $D$ to compute the frequency
of all the colours in $\Range(v)$ in $\Theta(|\Tree(v)|)$ time.  Let
$\NumCandidates$ be the value from Lemma \ref{lem:rebuildlists}.
Since there are at most $O(|\Tree(v)|)$ colours, we can use a linear
time selection algorithm to find the $\NumCandidates$-th most frequent
colour in $D$, and then find the top $\NumCandidates$ most frequent
colours via a linear scan in $O(|\Tree(v)|)$ time.  We can then
enforce the necessary semi-ordering on this list in $O(\NumCandidates)
= O(\frac{1}{\alpha})$ time, as described in
Section~\ref{sec:supquery}.  Thus, each leaf in $\Tree(v)$ pays $O(1)$
cost every $\Theta(|\Tree(v)|\alpha)$ insertions, or $O(1/\alpha)$
amortized cost per insertion.  Since each update may cause $O(\lg n)$
lists to be rebuilt, this increases the cost to $O((\lg n) / \alpha)$
amortized time per update.

We make use of standard local rebuilding techniques to keep the tree
$\Tree$ balanced, rebuilding the lists in nodes that are merged or
split during an update.  Since a node $v$ will only be merged or split
after $O(|\Tree(v)|)$ updates by the properties of weight-balanced
B-trees, these local rebuilding operations require $O(\lg n/\alpha)$
amortized time.  Finally, we can update $F_c$ and $F$ during an
insertion or deletion of a point of colour $c$ in $O(\lg n)$ time.
Thus, updates require $O((\lg n) / \alpha)$ amortized time overall,
and are dominated by the costs of maintaining the lists
$\CandidateList(v)$ in each node $v$. \qed
\end{proof}

\section{\label{sec:intkeys}Speedup for Integer Coordinates}
\newcommand{\findtop}{\textsc{Findtop}}
\newcommand{\LCA}{\textsc{LCA}}

We next describe how to improve the query time of the data structure
from Theorem~\ref{thm:dyn-1d-alpha-maj} from $O((\lg n) / \alpha)$ to
$O( \lg n / (\alpha \lg \lg n))$ for the case in which the
$x$-coordinates of the points in $\PointSet$ are integers that can be stored
in a constant number of words.

To accomplish this goal, we require an improved one-dimensional range
counting data structure, which we get by combining two existing data
structures.  The fusion tree of Fredman and Willard~\cite{Fredman1993}
is an $O(n)$ space data structure that supports predecessor and
successor queries in $O(\lg n / \lg \lg n)$ time and
insertions/deletions in $O(\lg n/ \lg \lg n)$ amortized time.  The
list indexing data structure of Dietz~\cite{Dietz1989} uses $O(n)$
space and supports \emph{rank queries}---i.e, given an element, return
the number of elements that precede it in the list--- in $O(\lg n/ \lg
\lg n)$ time, and insertions/deletions in $O(\lg n/ \lg \lg n)$
amortized time.  In Andersson~\etal~\cite{Andersson1999}, it was
observed that these data structures could be combined to support
dynamic one-dimensional range counting queries in $O(\lg n/ \lg \lg
n)$ time per operation; amortized for updates.  We refer to this data
structure as an \emph{augmented fusion tree}.

In order to achieve $O(\lg n/ (\alpha \lg \lg n))$ query time, we
implement all the range counting data structures as augmented fusion
trees: i.e., the data structures $F$, and $F_c$ for each $c \in C$.
Immediately, we get that we can perform a query in $O(\lg n / (\alpha
\lg \lg n) + \lg n)$ time: $O(\lg n / (\alpha \lg \lg n))$ time for
the range counting queries, and $O(\lg n)$ time to find the nodes in
$\CanonicalNodes_t$.  We now discuss how to remove the additive $O(\lg
n)$ term, which involves modifying our weight-balanced B-tree to
support dynamic lowest common ancestor queries.  To identify the top
$O(\lg{\frac{1}{\alpha}})$ levels of $\CanonicalNodes$, we use the
following lemma:

\begin{lemma}
\label{lem:lca}
The weight-balanced B-tree $\Tree$ can be augmented in order to support
lowest common ancestor queries in $O(\sqrt{\lg n})$ time without
changing the $O((\lg n/\alpha))$ amortized time required for updates.
\end{lemma}

\begin{proof}
Let the first ancestor of a node $u$ be the parent of $u$, and the
$\ell$-th ancestor of $u$ be the parent of the $(\ell -1)$-th ancestor
of $u$ for $\ell > 1$. In order to support lowest common ancestor
queries between two nodes $z_a$ and $z_b$, denoted $\LCA(z_a,z_b)$, we
add three pointers to each node $u \in \Tree$: pointers to both the
leaves representing the minimum and maximum $x$-coordinates in
$\Tree(u)$, and a pointer to the $\ell$-th ancestor of $u$; we will
fix the value of $\ell$ later. We can search for the $\LCA(z_a,z_b)$
by setting $v = z_a$ and following the pointer to the $\ell$-th
ancestor of $v$, denoted $v'$.  By checking the maximum $x$-coordinate
to see if $R(v')$ contains $z_b$, we can determine whether $v'$ is an
ancestor of $\LCA(z_a,z_b)$ or a descendant of $\LCA(z_a,z_b)$ in
constant time.  If $v'$ is a descendant of $\LCA(z_a,z_b)$, then we
set $v$ to $v'$ and $v'$ to the $\ell$-th ancestor of $v'$.  If $v'$
is an ancestor of $\LCA(z_a,z_b)$, then we backtrack and walk up the
path from $v$ to $v'$ until we find $\LCA(z_a,z_b)$.  Overall, it
takes $O(h_0 / \ell + \ell)$ time to find node $z = \LCA(z_a,z_b)$, if
$z$ is at height $h_0$ in $\Tree$.  By setting $\ell = O(\sqrt{\lg
  n})$ we get $O(\sqrt{\lg n})$ time.  Furthermore, the pointers we
added to $\Tree$ can be updated in $O(\lg n)$ amortized time during an
insertion or deletion.  Whenever we merge or split a node $u$, we have
$O(|\Tree(u)|)$ time to fix all of the pointers into $u$, by
properties of weight-balanced B-trees.  The pointers out of $u$ can be
fixed in $O(\lg n)$ worst case time. \qed
\end{proof}

Although Lemma~\ref{lem:lca} is weaker than other results
(cf.~\cite{Tsakalidis1988}), it is simple and sufficient for our
needs.  We next present the following theorem:

\begin{theorem}\label{thm:dyn-1d-alpha-maj-ints}
Given a set $\PointSet$ of $n$ points in one dimension with integer
coordinates and a fixed $\alpha \in (0,1)$, there is an $O(n)$ space
data structure that supports range $\alpha$-majority queries on $\PointSet$ in
$O(\lg n /(\alpha \lg \lg n))$ time, and both insertions and deletions
into $\PointSet$ in $O((\lg n) / \alpha))$ amortized time.
\end{theorem}

\begin{proof}
Suppose we are given a query range $[x_a,x_b]$.  Applying
Lemma~\ref{lem:lca} to the weight-balanced B-tree $\Tree$, we claim
that we can identify the top $\ell$ levels of $\CanonicalNodes$-- that
are \emph{not} necessarily from consecutive levels in $\Tree$-- using
$O(\ell)$ least common ancestor operations.  To show this, we describe
a recursive procedure $\findtop(z_a, z_b, \ell)$ for identifying the
top $\ell$ levels of $\CanonicalNodes$.  We assume that we have
acquired pointers to $z_a$ and $z_b$, the leaves of $\Tree$ that
represent the $x$-coordinates of the successor of $x_a$ and
predecessor of $x_b$, respectively.  To do this, we add a pointer from
each leaf in the augmented fusion tree $F$ to its corresponding leaf
in $\Tree$.  Given a query, we initially perform a successor query for
$x_a$ and predecessor query for $x_b$ in $F$, and follow these extra
pointers to $z_a$ and $z_b$, respectively.  We assume that $z_a \neq
z_b$, otherwise the query is trivially answered by reporting the
colour stored in $z_a$.

Let $z = \LCA(z_a,z_b)$, and $c_i$ denote the $i$-th child of $z$.
Let $z_l$ and $z_r$ denote the leftmost and rightmost leaves in $\Tree_z$.
In constant time we can determine children $c_j$ and $c_k$ of $z$
which are on the path to $z_a$ and $z_b$, respectively.  Note that $k
- j > 0$, otherwise $z$ is not the $\LCA(z_a,z_b)$.  We say we are in
the \emph{good case} when $z_a = z_l$, $z_b = z_r$, and/or $k - j >
1$.  When we are in the good case, either $c_j$, $c_k$, and/or
$c_{j+1}, ..., c_{k-1}$ are in the top level of $\CanonicalNodes$, and we set $\ell'
= \ell - 1$.  Otherwise, if $k - j = 1$ and $z_a \neq z_l$ and $z_b
\neq z_r$, then we are in the \emph{bad case}.  In the bad case we
have not found the top level of $\CanonicalNodes$, and we set $\ell' = \ell$.  In
both cases (good or bad), let $z_{b'}$ be the leaf in $c_k$
representing the minimum $x$-coordinate in $\Tree_{c_k}$, and $z_{a'}$ be
the leaf in $c_j$ representing the maximum $x$-coordinate in
$\Tree_{c_j}$.  We recurse if $\ell' > 0$, calling $\findtop(z_a, z_{a'},
\ell')$ if $z_a \neq z_{a'}$ and $\findtop(z_{b'},z_b,\ell')$ if $z_b
\neq z_{b'}$.

We observe that the procedure $\findtop(z_{a}, z_b, \ell)$ uses
$O(\ell)$ least common ancestor queries.  This is because if a call to
$\findtop$ is in the bad case, then the subsequent recursive call(s)
will be in the good case by choice of $z_{a'}$ and $z_{b'}$, and only
the initial call to $\findtop$ can make two recursive calls.  Using
$\findtop$, we can identify the top $O(\lg{\frac{1}{\alpha}})$ levels
of $\CanonicalNodes$ in $O(\sqrt{\lg n} \times \lg{\frac{1}{\alpha}})$ time,
replacing the $O(\lg n)$ additive term.  This factor is strictly
asymptotically less than the time required to perform the
range-counting queries, which is $O(\lg n/(\alpha \lg \lg n))$.  

By Lemma~\ref{lem:lca}, the we can support the lowest common ancestor
operation without increasing the update time of $\Tree$ as stated in
Theorem~\ref{thm:dyn-1d-alpha-maj}.  The extra pointers we added from
the leaves of $F$ to the leaves of $\Tree$ can also be updated without
affecting the bound from Theorem~\ref{thm:dyn-1d-alpha-maj}, since
during any insertion/deletion of a point $p$, the two leaves
corresponding to $p$ in both $F$ and $\Tree$ must be located.  Therefore,
the total update time follows from
Theorem~\ref{thm:dyn-1d-alpha-maj}. \qed
\end{proof}

% Pat: Edited to here

\section{\label{sec:higher-d}Extension to Higher Dimensional Point Sets}

In this section we present a refinement of the technique presented by
Karpinski and Nekrich~\cite{Karpinski2008}, who used standard range
tree techniques~\cite{Bentley1980} to generalize their range
$\alpha$-majority structures to higher dimensions.\footnote{In the
  preliminary version of this paper~\cite{HMN11}, the paragraph
  preceding Theorem~3 erroneously stated that the result which we
  prove in this section follows immediately from the analysis of
  Karpinski and Nekrich~\cite{Karpinski2008}.}  We note that,
recently, Wilkinson~\cite{W12} has used the same refinement to improve
the bounds of Durocher~\etal~\cite{DHMNS2012} for the two-dimensional
static case.

All of the algorithms presented thus far have the following two phase
structure.  The first is the \emph{candidate extraction} phase, in
which we extract a list of candidates from our data structure.  The
second phase is the \emph{verification} phase, in which we use range
counting data structures to verify that they are actual
$\alpha$-majorities.  For higher dimensional problems we speed up the
verification phase by adding an additional filtering phase between the
candidate extraction and verification phases.

In order to do this, we make use of \emph{approximate range counting}
data structures~\cite{M06,Nekrich2009b,W12}. If $\NumPointsInRange$
points are contained in the query range, then an approximate range
counting data structure with additive error $\NumPointsInRange'$ will
return a count in the range $[\NumPointsInRange -
  \NumPointsInRange',\NumPointsInRange +\NumPointsInRange']$;
see~\cite{Nekrich2009b}.  Similarly, a data structure with
multiplicative error $(1-\varepsilon)$ will return a count in the
range $[(1-\varepsilon)\NumPointsInRange,\NumPointsInRange]$;
see~\cite{M06}.  In the remainder of this section we first modify
existing data structures for approximate range counting, and then
consider their applications to higher dimensional data structures for
dynamic range $\alpha$-majority queries.

\subsection{Approximate Range Counting}

Before stating the results for higher dimensional range
$\alpha$-majority data structures we require some additional results
on approximate range counting.  We begin with a lemma, which is a very
minor generalization of Nekrich's one-dimensional approximate range
counting data structure~\cite[Theorem 1]{Nekrich2009b}.  In the
original structure each point is unweighted, but we wish to add the
operations $\texttt{increment}$ and $\texttt{decrement}$ to the
structure, which respectively increase and decrease the weight of a
point by one.  We assume a newly inserted point begins with weight
one.  Instead of returning the number of points in a query range
(within an additive error term), our query operation will return the
sum of the weights of the points in the query range, within an
additive error term.

\begin{lemma}\label{lem:wapprox}
Let $\tau > 1$ be an integer constant, $\texttt{dpred}(n)$ denote the
cost of a dynamic predecessor search on $n$ keys, and
$\NumPointsInRange$ denote the sum of the weight of the points
contained the query range.  There exists an $O(n)$ space data
structure that supports \emph{approximate weighted range counting
  queries} with additive error $\NumPointsInRange^{1/\tau}$ in
$O(\texttt{dpred}(n) + \lg \lg n)$ time, deletions in $O(\lg \lg n)$
amortized time, and insertions in $O(\texttt{dpred}(n) + \lg \lg n)$
amortized time.  The operations $\texttt{increment}$ and
$\texttt{decrement}$ are supported in $O(\lg \lg n)$ amortized time.
\end{lemma}

\begin{proof}
Let $\NumPointsInRange'$ be the approximate weight returned by our
data structure, while $\NumPointsInRange$ is the exact weight.  We
divide the solution into two cases. In the first case, we assume that
$\NumPointsInRange \ge h_0(\lg n)^\tau$ for some arbitrary constant
$h_0 > 0$.  We emphasize that in both cases the data structure and
proof are essentially the same as Theorem 1 of
Nekrich~\cite{Nekrich2009b}, with some minor modifications.  We use an
exponential search tree $\Tree$~\cite{AT07}, where each leaf in
$\Tree$ represents a point, but also stores the weight associated with
the point. We require some additional notation, and closely follow
that of Nekrich~\cite{Nekrich2009b}. Let $v_i$ denote the $i$-th child
of $v$, $n_v$ denote the number of leaves in the subtree $\Tree(v)$,
$\WeightFunct(v)$ denote the weight of the leaves of the subtree
$\Tree(v)$, and $\NodeDegree(v)$ denote the number of children of $v$.
In the exponential search tree $\Tree$, each node $v$ has
$\Theta(n_v^{1/\tau})$ children, each of which contains between
$n_v^{(\tau-1)/\tau}/2$ and $2n_v^{(\tau-1)/\tau}$ points, for a fixed
constant $\tau > 2$. Each node $v$ stores its weight
$\WeightFunct(v)$, as well as a set of approximate weights
$\WeightFunct'(v,i,j)$, such that

\begin{equation}\notag
\WeightFunct(v_i) + ... + \WeightFunct(v_j) - n_v^{3/\tau}/2 \le \WeightFunct'(v,i,j) \le \WeightFunct(v_i) +... + \WeightFunct(v_j) + n_v^{3/\tau}/2 \enspace ,
\end{equation}

\noindent
for all $1 \le i \le j \le \NodeDegree(v)$.  We recompute all counts
$\WeightFunct'(v,i,j)$ after $n_v^{3/\tau}/2$ update operations
(insertions, deletions, increments, and decrements).  Recomputing all
the $\WeightFunct'(v,i,j)$'s for a node takes $O(n^{2/\tau}_v)$
time. Thus, each update operation--- insertion, deletion, increment,
decrement--- requires $O(\lg \lg n)$ amortized time, since the height
of $\Tree$ is $O(\lg \lg n)$ and we must increment or decrement the
weight $\WeightFunct(v)$ stored in each node on the path from leaf to
root.

The space is linear by the properties of exponential search
trees~\cite{AT07}, and all that remains is to argue the correctness of
the query algorithm of Nekrich~\cite{Nekrich2009b}.  The query
algorithm essentially finds the ranges in $\Tree$ that represent the
query range, and returns the summation of the approximate counters of
those ranges.

For a fixed node $v$ from the set of nodes representing the query
range, with children $v_i, ..., v_j$ contained entirely in the query
range, let $\NumPointsInRange'_v = \WeightFunct'(v,i,j)$ and
$\NumPointsInRange_v = \sum_{\ell=i}^{j} \WeightFunct(v_\ell)$.  Then
$\NumPointsInRange_v - n_v^{3/\tau} \le \NumPointsInRange'_v \le
\NumPointsInRange_v + n_v^{3/\tau}$.  Since $\NumPointsInRange_v \ge
n_v^{(\tau-1)/\tau}$, $\NumPointsInRange_v -
\NumPointsInRange_v^{3/(\tau-1)} \le \NumPointsInRange'_v \le
\NumPointsInRange_v + \NumPointsInRange_v^{3/(\tau-1)}$.  Since
$\Tree$ has height $h_1 \lg \lg n$ for some constant $h_1$,
$\NumPointsInRange - (2h_1\lg \lg n) (\NumPointsInRange^{3/(\tau-1))})
\le \NumPointsInRange' \le \NumPointsInRange+(2h_1 \lg \lg
n)(\NumPointsInRange^{3/(\tau-1)})$.  However, since we assume
$\NumPointsInRange \ge h_0 (\lg n)^\tau$, we need only ensure $h_0 >
(2h_1)^\tau$ in order for $2h_1 \lg \lg n \le
\NumPointsInRange^{1/(\tau-1)}$.  Thus, $\NumPointsInRange -
\NumPointsInRange^{4/(\tau-1))} \le \NumPointsInRange' \le
\NumPointsInRange + \NumPointsInRange^{4/(\tau-1)}$.  By replacing
$\tau$ with $\tau' = \max(5\tau,5)$ we obtain the result of the lemma.

In the second case, when $\NumPointsInRange < h_0(\lg n)^\tau$, we
make use of an alternative data structure.  We divide the point set
into groups of between $h_0(\lg n)^\tau$ and $4h_0(\lg n)^\tau$
consecutive points, and store each group in a balanced binary search
tree.  Each node $u$ in the search tree stores the total weight of the
nodes in the subtree induced by $u$.  Given the successor and
predecessor, $e_1'$ and $e_2'$, of a query range $[e_1,e_2]$, we can
assume that points $e_1'$ and $e_2'$ either belong to the same group,
or two adjacent groups.  Thus, given $e_1'$ and $e_2'$, which can be
obtained in $O(\texttt{dpred}(n))$ time, we can tally the \emph{exact}
weight in this case in $O(\lg \lg n)$ time.  Using standard techniques
we can support insertion in $O(\texttt{dpred}(n) + \lg \lg n)$
amortized time; $O(\texttt{dpred}(n))$ to find the position in which
to insert the new element, and $O(\lg \lg n)$ amortized time to insert
it into the binary search tree for its group, accounting for
merging/splitting of groups.  By analogous arguments deletion takes
$O(\lg \lg n)$ amortized time.  Finally, increment and decrement can
be performed in $O(\lg \lg n)$ worst case time. \qed
\end{proof}

Before continuing, we require the definition of a generalized
union-split-find (GUSF) data structure, as well as the time bounds for
its operations.

\begin{lemma}[\cite{GK2009}, Theorem 5.2]\label{lem:gusf}
A GUSF stores an ordered list $G$ of elements, in which each element
$x$ of $G$ is associated with a subset $\GUSFColours(x) \subseteq \{
1, ..., \lg^{\frac{1}{4}} n \}$ of colours.  Assume we have a pointer
to an element $x \in G$, and $\GUSFSubsetColours \subseteq \{1,
... ,\lg^{\frac{1}{4}} n\}$ be a set of colours.  A GUSF supports the
operations:
\begin{itemize}
\item $\texttt{find}(x,\GUSFSubsetColours)$: return the successor of
  $x$ with colour $c \in \GUSFSubsetColours$.
\item $\texttt{add}(y,x)$: inserts $y$ into the list before $x$.
\item $\texttt{erase}(x)$: removes $x$ from the list, assuming
  $\GUSFColours(x) = \emptyset$.
\item $\texttt{mark}(x,c)$: inserts $c$ into $\GUSFColours(x)$.
\item $\texttt{unmark}(x,c)$ removes $c$ from $\GUSFColours(x)$.
\end{itemize}
\noindent
A GUSF can be implemented in $O(n)$ space, such that each operation
takes $O(\lg \lg n)$ time.  The time bound for $\texttt{add}$ and
$\texttt{erase}$ is amortized, while the running time of all other
operations are worst case.  A GUSF containing $n$ elements can be
constructed in $O(n \lg^{\frac{1}{8}}n \lg \lg n)$ time.
\end{lemma}

Next we are ready to state and prove the main result of this section:

\begin{lemma}\label{lem:approx}
Let $\PointSet$ be a set of $d$-dimensional points for any $d \ge 2$.  The
point set, $\PointSet$, can be preprocessed into an $O((n \lg^{d-1}n)/ \lg \lg
n)$ space data structure, such that for any arbitrary $d$-dimensional
axis-aligned hyperrectangle, $\QueryRange$, approximate range counting queries
can be performed in $O(\lg^{d-1} n)$ time, with additive error $|\PointSet
\cap \QueryRange|^{\frac{1}{j}}$, for any constant integer $j > 1$.  Insertions
and deletions can be performed in $O(\lg^{d-1 +\frac{1}{4}} n)$
amortized time.
\end{lemma}

\begin{proof}
The proof of this lemma is \emph{almost} the same as Theorem 2
in~\cite{Nekrich2009b}, except that we increase the cost of the query
by a factor of $O(\lg \lg n)$ for any $d \ge 3$, and decrease the
space bound by a factor of $O(\lg^{\varepsilon}n)$. We make use of
dynamic fractional cascading in weight balanced B-trees without
modifications~\cite{GK2009}, and a slight modification of the GUSF of
Lemma~\ref{lem:gusf}.

We next describe how to combine the one-dimensional weighted
approximate counting data structure from Lemma~\ref{lem:wapprox} with
the GUSF.  This will allow the GUSF to support coloured approximate
range counting: i.e., given a colour $c \in \{ 1, ...,
\lg^{\frac{1}{4}} n \}$ and a range $[x_a,x_b]$, approximately report
the number of elements with colour $c$ contained in $[x_a,x_b]$.  A
single \emph{modified} GUSF will then be stored in each internal node
of a weight-balanced B-tree, in order to support two-dimensional
approximate range counting.

A GUSF groups consecutive elements into \emph{blocks} which are of
size $\Theta(\lg^{2+\frac{1}{4}} n )$.  The elements in each block are
stored in a balanced binary search tree.  For each node in the tree,
we store the counts of the number of children with each of the
$\lg^{\frac{1}{4}}n$ colours, with counter $n_c$ storing the number of
points of colour $c$.  Since the tree has $O(\lg^{2+\frac{1}{4}} n )$
elements, each counter requires $O(\lg \lg n)$ bits, and thus the
counters for a node can be packed into a constant number of words.
Thus, these counters do not increase the space of the GUSF structure
asymptotically.

As in a standard GUSF, each block in the modified GUSF is represented
in an order maintenance structure that maps a block to an integer
coordinate.  Given two blocks, $b$ and $b'$, we denote their
corresponding integer coordinates $X(b)$ and $X(b')$, and we can
determine whether the elements in $b$ precede those in $b'$, or vice
versa, by comparing these coordinates; see~\cite{GK2009} for full
details.

Our modified GUSF also stores $O(\lg^{\frac{1}{4}} n)$ copies of the
data structure from Lemma~\ref{lem:wapprox}, one for each colour $c
\in [1, \lg^{\frac{1}{4}} n]$, denoted $D_c$.  We discuss how to set
the parameter $\tau$ for each $D_c$ later.  For each block $b$ that
has a counter value $n_c > 0$ in its root for colour $c$, we store a
point representing that block in $D_c$, with weight $n_c$, and
coordinate $X(b)$.  The root of $b$ also stores a pointer to the leaf
in $D_c$ representing these points, for each colour $c \in [1,
  \lg^{\frac{1}{4}} n]$.  Since there are at most $O(n / \lg^{2 +
  \frac{1}{4}} n)$ blocks, all these structures together occupy
$O(n / \lg^{2} n)$ space.

Given two elements $e_1$ and $e_2$, where $e_1 < e_2$ and both
elements are marked with colour $c$, we can determine the approximate
number of points with colour $c$ that both succeed $e_1$ and precede
$e_2$, as follows.  First, if $e_1$ and $e_2$ are in the same block,
we can return the \emph{exact count} in $O(\lg \lg n)$ time using the
counters that are stored in the nodes of the balanced binary tree
representing the block.  Otherwise, we need to perform an additional
step of querying the data structure $D_c$, providing pointers to the
leaves in $D_c$ that represent the blocks containing $e_1$ and $e_2$,
respectively.

With the exception of the data structures $D_c$, the GUSF containing
$n$ elements can be constructed in $O(n \lg^{\frac{1}{8}}n \lg \lg n)$
time by Lemma~\ref{lem:gusf}, since each GUSF operation takes at most
$O(\lg \lg n)$ amortized time.  Since each point results in an
insertion or increment operation on $O(\lg^{\frac{1}{8}}n)$
approximate range counting data structures, each of size $O(n /
\lg^{2+\frac{1}{4}} n)$, this takes $O(\lg^{\frac{1}{8}}n \lg \lg n)$
amortized time per point, and does not asymptotically change the
construction time.

We are next ready to discuss our data structure for planar approximate
counting, i.e., the case in which $d=2$.  We store a weight balanced
B-tree $\Tree$ over the $y$-coordinates of the given points, with
branching parameter $\Theta(\lg^{\frac{1}{8}} n)$ and leaf parameter
$1$.  For each internal node $u$ of $\Tree$ with degree $f$, we store
our modified GUSF $M(u)$, over all of the points in the subtree
$\Tree(u)$, ordered by their $x$-coordinate.  Note that there are
$\Theta(\lg^{\frac{1}{4}}n)$ possible contiguous subranges of children
of $u$ in total, and each child of $u$ belongs to
$\Theta(\lg^{\frac{1}{8}} n)$ of these ranges $[i,j]$, where $1 \le i
\le j \le f$.  We construct a set of colours, and each colour
corresponds to a possible range $[i,j]$. Thus, each point in $M(u)$ is
marked with the $\Theta(\lg^{\frac{1}{8}}n)$ colours corresponding to
these ranges.  Each node $u$ also stores a catalogue $V(u)$
corresponding to the points in $\Tree(u)$ ordered by $x$-coordinate.
Each catalogue element stores a pointer to the corresponding element
in $M(u)$. We maintain a dynamic fractional cascading data structure
over the catalogues of $\Tree$.

Since the branching parameter of the tree $\Tree$ is
$\Theta(\lg^{\frac{1}{8}}n)$, the tree has height $\Theta(\lg n / \lg \lg n)$.
Each point is stored in $\Theta(\lg n / \lg \lg n)$ nodes, each containing
a constant number of linear space structures. Thus, the space occupied
by the data structure is $\Theta(n \lg n / \lg \lg n)$.

To answer a query of the form $[x_1, x_2] \times [y_1,y_2]$, we
perform a search for the successor and predecessor, $e_1$ and $e_2$,
of the query range $[x_1,x_2]$ in each catalogue of each node among
the nodes representing $[y_1,y_2]$ in $\Tree$.  This takes $\Theta(\lg
n)$ time, since there are $\Theta(\lg n / \lg \lg n)$ catalogues: the
initial search requires $\Theta(\lg n)$ time, and each additional
search uses $\Theta(\lg \lg n)$ time.  For a fixed internal node $u$
of $\Tree$, such that the query range $[y_1, y_2]$ spans children
$[i,j]$, let $c$ be the colour representing $[i,j]$ in $M(u)$.  We
locate $e_1'$ and $e_2'$, the respective successor and predecessor of
$e_1$ and $e_2$ in $M(u)$ with colour $c$, using the $\texttt{find}$
operation.  Thus, locating $e_1'$ and $e_2'$ in $M(u)$ takes $O(\lg
\lg n)$ time.  Once we have located $e_1'$ and $e_2'$ we can query
$D_c$, and the counters in the block(s) containing $e_1'$ and $e_2'$,
in $O(\lg \lg n)$ time, as outlined above.  Thus, the overall query
time is $O(\lg \lg n (\lg n / \lg \lg n) + \lg n )$ which is $O(\lg
n)$.

Suppose we desire additive error $|\PointSet\cap
\QueryRange|^{\frac{1}{j}}$ for some fixed constant $j > 1$.  Then, we
set the parameter $\tau = 2j$.  Let $\NumPointsInRange'$ denote the
sum of each of the $h_2(\lg n / \lg \lg n)$ approximate counts tallied
at each node that represents the query range, where $h_2$ is a
constant that depends on the height of $\Tree$, and
$\NumPointsInRange$ denotes the exact count.  Thus,
\begin{equation}
\NumPointsInRange - h_2(\lg n / \lg \lg n) \NumPointsInRange^{\frac{1}{\tau}} \le \NumPointsInRange' \le \NumPointsInRange + h_2(\lg n /
\lg \lg n)\NumPointsInRange^{\frac{1}{\tau}} \enspace.
\end{equation}

If $\NumPointsInRange \ge (h_2 \lg n)^\tau$, then
$\NumPointsInRange^{\frac{1}{\tau}} \ge h_2(\lg n / \lg \lg n)$.
Thus, $\NumPointsInRange - \NumPointsInRange^{\frac{2}{\tau}} \le
\NumPointsInRange ' \le \NumPointsInRange +
\NumPointsInRange^{\frac{2}{\tau}}$.  Since $\tau = 2j$, we are left
with $\NumPointsInRange - \NumPointsInRange^{\frac{1}{j}} \le
\NumPointsInRange' \le \NumPointsInRange +
\NumPointsInRange^{\frac{1}{j}}$, which is the desired error term.

Next suppose $\NumPointsInRange <(h_2\lg n)^\tau$.  In this case, we
can retrieve the exact count in $O(\lg n)$ time using the binary tree
representation of $D_c$, since none of the ranges represented can
contain more than $\NumPointsInRange$ points.  Thus, in both cases we
have shown the query algorithm is correct.  Note that we must ensure
$h_0 \ge h_2^\tau$, in addition to the constraints on $h_0$ described
in Lemma~\ref{lem:wapprox}.

In order to insert a point $p$, we identify the nodes on the path $Y$
from the root of $\Tree$ to the leaf where $p$ will be inserted.  We
then search for the successors of $p$ in all of the catalogues on this
path, which takes $O(\lg n)$ time in total.  Once we have the
successor, we can insert $p$ into each GUSF along $Y$ in the following
way.  Let $u$ be a node in $Y$ and $u_i$ be the child of $u$ whose
range contains $p$.  Using the pointer to the successor of $p$ in
$M(u)$, we can perform an \texttt{add} operation, inserting $p$ into a
block $b$ in $M(u)$.  Let $\GUSFSubsetColours$ denote the set of
colours in $M(u)$ representing the ranges that contain $u_i$.

If $b$ splits into two blocks $b$ and $b'$ as a result of this, then
we must decrement the weight of the element representing $b$ in each
$D_c$ for each $c$ with a non zero counter in the root of $b$.  We
also must insert a new element representing $b'$ into each $D_c$ for
each $c$ with a non zero counter in the root of $b'$, and increment
its weight accordingly.  Recall that $O(\lg^{2+\frac{1}{4}}n)$
elements must have been inserted into $M(u)$ to cause $b$ to split.
Each split causes $O(\lg^{\frac{1}{8}}n (\lg^{2+\frac{1}{4}} n))$
update operations on all $D_c$, for each $c$ stored in the roots of
$b$ and $b'$.  This is $O(\lg^{\frac{1}{8}}n \lg \lg n)$ amortized
update time.  In the case in which $b$ does not split, we still
require $O(\lg^{\frac{1}{8}}n \lg \lg n)$ amortized time to
\texttt{increase} the weight of the representative of $b$ in each
$D_c$, for $c \in \GUSFSubsetColours$.  Since there are $O(\lg n / \lg
\lg n)$ nodes in $Y$, the overall insertion time is thus
$O(\lg^{1+\frac{1}{8}}n)$, provided we do not cause a node to split or
two nodes to merge in the base tree $\Tree$.  Deletion is handled
analogously, except that we \texttt{decrease} the weight of the
representative of $b$ in each $D_c$ for $c \in \GUSFSubsetColours$.
Thus, deletion requires $O(\lg^{1+\frac{1}{8}}n)$ amortized time as
well.

In the case in which a node $u \in \Tree$ splits or merges, we can efficiently
update the fractional cascading data structure using the techniques
described in~\cite{GK2009}.  The cost of a split or merge is dominated
by the cost of rebuilding the modified GUSF structures in both $u$ and
$u$'s parent.  We can rebuild each modified GUSF in a node
representing $m$ points in $O(m \lg^{\frac{1}{8}}n \lg \lg n)$ time.
Since $O(m)$ updates are required to split a node with a parent
containing $O(m \lg^{\frac{1}{8}} n)$ points, and $O(\lg n / \lg \lg
n)$ splits/merges can occur during a single update, the cost of
performing an update is $O(\lg^{1 + \frac{1}{4}}n)$ amortized time.

To get the bound stated by the lemma for higher dimensions, we use the
standard range tree technique~\cite{Bentley1980}, which inflates the
space, query and update time by a factor of $O(\lg n)$ for each
additional dimension.  In general, we must set the parameter $\tau =
2^{d-1}j$.  \qed
\end{proof}

\subsection{Range $\alpha$-Majority in Higher Dimensions}
As an application of Theorem~\ref{thm:dyn-1d-alpha-maj}, and the
approximate range counting data structures of Lemma~\ref{lem:approx},
we can improve the query time for range $\alpha$-majority data
structures in higher dimensions.

\begin{theorem}\label{thm:dyn-ddim-alpha-maj}
Given a set $\PointSet$ of $n$ points in $d$ dimensions (for a constant $d \ge
2$) and a fixed $\alpha \in (0,1)$, there is an $O(n\lg^{d-1} n)$
space data structure that supports range $\alpha$-majority queries on
$\PointSet$ in $O((\lg^{d} n) / \alpha)$ time, and insertions and deletions
into $\PointSet$ in $O((\lg^{d} n)/ \alpha)$ amortized time.  
\end{theorem}

\begin{proof}
Using range trees, we can convert any $d$-dimensional range
$\alpha$-majority query into a combination of several
$(d-1)$-dimensional range $\alpha$-majority queries and
$d$-dimensional range counting queries.  In particular, let
$\mathcal{S}(n,d)$ denote the cost of a $d$-dimensional range counting
query on a dynamic set of $n$ points, and $\mathcal{M}(n,d)$ denote
the cost of a $d$-dimensional range $\alpha$-majority on a dynamic set
of $n$ points.  Then, for $d \ge 2$,

\begin{equation}\label{eq:recurr1}
\mathcal{M}(n,d) = O(\lg n)\mathcal{M}(n,d-1) + O(\lg n /\alpha)\mathcal{S}(n,d) \enspace , 
\end{equation}

\noindent 
since we can extract and verify the frequency of the $O((\lg n)
/\alpha)$ candidates from the $O(\lg n)$ nodes representing the range
spanned by the $d$-th coordinate of the query range.  Note that each
candidate is an $\alpha$-majority if we consider their first $(d-1)$
coordinates only.  Since $d$-dimensional dynamic orthogonal range
counting queries require $O(\lg^d n)$ time~\cite{Chazelle1988},
$\mathcal{M}(n,d) = O((\lg^{d+1} n)/ \alpha)$.

To further reduce the query time we observe that only $O(1/\alpha)$ of
the $O(\lg n / \alpha)$ candidates can have frequency above
$(1-\varepsilon)\alpha q$, where $q$ is the total number of points
contained in the query range and $\varepsilon \in (0,1)$ is an
arbitrary constant.  Thus, we add additional data structures $F'_c$
for each $c \in \ColourSet$, where each $F'_c$ is the structure of
Lemma~\ref{lem:approx} and stores the points of colour
$c$.\footnote{The data structure of Lemma~\ref{lem:approx} has very
  small additive error, though, for the purposes of this proof, we
  need only constant multiplicative error.}  Using these data
structures, we can perform an additional filtering pass of the list of
$O(\lg n / \alpha)$ candidates into a shorter list of $O(1/\alpha)$
candidates. After this filtering step we can then verify the frequency
of the $O(1/\alpha)$ candidates above this threshold exactly using
range counting data structures $F_c$.  Let $\hat{S}(n,d)$ denote the
query time of the data structure from Lemma~\ref{lem:approx}.  We can
rewrite the recurrence of Equation~\ref{eq:recurr1} as:

\begin{equation}\label{eq:recurr2}
\mathcal{M}(n,d) = O(\lg n) \mathcal{M}(n,d-1) + O(\lg n / \alpha) \hat{S}(n,d) + O(1/\alpha) \mathcal{S}(n,d) \enspace .
\end{equation}

\noindent 
This recurrence resolves to $\mathcal{M}(n,d) = O((\lg^d n)
/\alpha)$. We can update the structures $F_c$ and $F'_c$ for the
colour, $c$, of the inserted or deleted point, and $F$ in $O(\lg^d n +
\lg^{d-1+\frac{1}{4}}n)$ amortized time. Each of the $O(\lg n)$
$(d-1)$-dimensional range $\alpha$-majority data structures can be
updated in $O((\lg^{d-1}) n/\alpha)$ amortized time, for a total of
$O((\lg^{d} n) / \alpha)$ amortized time.  Finally, the space cost is
dominated by the range $\alpha$-majority structures.  The space
occupied by this can be expressed as $\mathcal{U}(n,d) = O(\lg
n)\mathcal{U}(n,d-1)$, where $\mathcal{U}(n,d)$ is the space occupied
by a $d$-dimensional dynamic range $\alpha$-majority structure, and $d
\ge 2$.  Thus $\mathcal{U}(n,d) = O(n \lg^{d-1} n)$.  \qed
\end{proof}

\section{Dynamic Arrays\label{sec:array}}

In this section we extend our results to dynamic arrays.  In the
dynamic array version of the problem, we wish to support the following
operations on an array $A$ of length $n$, where each $A[i]$ contains a
colour, for $1 \le i \le n$:

\begin{itemize}

\item $\textsc{Insert}(i,c)$: Insert the colour $c$ between the
  colours $A[i-1]$ and $A[i]$.  This shifts the colours in positions
  $i$ to $n$ to positions $i+1$ to $n+1$, respectively.

\item $\textsc{Delete}(i)$: Delete the colour $A[i]$.  This shifts the
  colours in positions $i+1$ to $n$ to positions $i$ to $n-1$,
  respectively.

\item $\textsc{Modify}(i,c)$: Set the colour $A[i]$ to $c$.

\item $\textsc{Query}(i,j)$: Let $|A[i..j]|_c$ denote the number of
  occurrences of colour $c$ in the range $A[i..j]$.  Report the set of
  colours $M$ such that for each $c \in M$, $|A[i..j]|_c > \alpha |j -
  i + 1|$.  As before, we refer to a colour $c \in M$ as an
  \emph{$\alpha$-majority} in the range $A[i..j]$, and the query as a
  \emph{range $\alpha$-majority query}.

\end{itemize}

The dynamic array problem boils down to the well-studied problem of
maintaining an injective order preserving mapping from the positions
in $A$ into a larger set of integer keys $\PointSet$
\cite{Nekrich2007}.  We next prove the following theorem:

\begin{theorem}\label{thm:dyn-array}
Given an array $A[1..n]$ of colours and a fixed $\alpha \in (0,1)$,
there is an $O(n)$ space data structure that supports range
$\alpha$-majority queries on $A$ in $O((\lg n) /(\alpha \lg \lg n))$
time, \textsc{Insert} in $O((\lg^3n) / (\alpha \lg \lg n))$ amortized
time, \textsc{Delete} in $O((\lg n) / \alpha)$ amortized time, and
\textsc{Modify} in $O((\lg n) / \alpha)$ amortized time.
\end{theorem}

\begin{proof}
We maintain our data structure $\Tree$ from
Theorem~\ref{thm:dyn-1d-alpha-maj-ints} on the integer key set
$\PointSet$.  Each time a key $p$ in $\PointSet$ is changed to key
$p'$, we must delete $p$ from $\Tree$, and then insert $p'$ into
$\Tree$.  If an insertion or deletion into our dynamic array changes
$\ell$ keys in the mapping, it will require $O((\ell\lg n) /\alpha)$
amortized time to change these values in $\Tree$. We note that a
\textsc{Modify} operation corresponds to one deletion and one
insertion into $\Tree$, requiring $O((\lg n) /\alpha)$ time.

We apply the \emph{dynamic reduction to extended rank space}
technique~\cite{Nekrich2007}, which maps the positions in $A$ to
integers in the bounded universe $[1..O(n^3)]$.  This mapping requires
$O((\lg^2n)/\lg\lg n)$ amortized time for insertions, and $O(1)$
amortized time for deletions.  These time bounds also bound the number
of key changes for insertion and deletion (in the amortized sense),
completing the proof. \qed
\end{proof}

\section{Conclusions}

We have presented several new dynamic data structures for the range
$\alpha$-majority problem.  These data structures improve on the
previous results in terms of space, query, and update time.  

Notably, for one-dimensional points, we presented a linear space data
structure with $O(\lg n / \alpha)$ query time, and $O(\lg n / \alpha)$
amortized update time.  In the case in which the coordinates of the
points are integers, we reduced the query time by a $(\lg \lg
n)$-factor.  This improved query time matches an existing cell-probe
lower bound, for the case when $1/\alpha$ is a constant, and the word
size is $\Theta(\lg n)$.

We also extended our one-dimensional data structure to $d$-dimensions,
where $d \ge 2$ is an arbitrary constant.  The generalized structure
occupies $O(n \lg^{d-1} n)$ space, has $O(\lg^{d} n / \alpha)$ query
time, and supports updates in $O(\lg^{d} n / \alpha)$ amortized time.
It would be interested to determine if either the space or query time
can be improved for the higher dimensional data structure.

\bibliographystyle{splncs03} 
\bibliography{dyn_range_maj}

\end{document}